\documentstyle[aps,multicol]{revtex}
\def\be{\begin{eqnarray}}
\def\ee{\end{eqnarray}}
\def\l{\langle}
\def\r{\rangle}

\def\state{{\cal S}}
\def\Cx{{\bf C}}
\def\Hi{{\cal H}}
\def\tr{{\rm Tr}}
\def\upN{^{\otimes N}}

\begin{document}
\title{Optimal Manipulations with Qubits: 
Universal NOT Gate
}
\author{
V. Bu\v{z}ek$^{1}$, M. Hillery$^{2}$, and R.F. Werner$^{3}$
}
\address{
$^{1}$
Institute of Physics, Slovak Academy of Sciences, D\'ubravsk\'a cesta 9,
842 28 Bratislava, Slovakia\newline
$^{2}$Department
of Physics and Astronomy, Hunter College, CUNY,
695 Park Avenue, New York, NY 10021, USA\newline
Inst. f. Mathematische Physik, TU Braunschweig, Mendelssohnstr. 3,
38304 Braunschweig, Germany
}

\date{15 January, 1999}
\maketitle
\begin{abstract}
It is not a problem to complement a classical bit, i.e. to change
the value of a bit, a $0$ to a $1$ and vice versa.  This is
accomplished by a NOT gate. Complementing a qubit in an {\it
unknown} state, however, is another matter. We show that this
operation cannot be done perfectly.
We define the Universal-NOT (U-NOT) gate which out of $N$
identically prepared pure input qubits generates $M$ output qubits
in a state which is as close as possible to the perfect
complement. This gate can be realized by classical estimation and
subsequent re-preparation of complements of the estimated state.
Its fidelity is therefore equal to the fidelity ${\cal F}=
(N+1)/(N+2)$ of optimal estimation, and does not depend on the
required number of outputs.
We also show that when some additional  {\it a priori} information
about the state of input qubit is available, than the fidelity of
the quantum NOT gate can be much better than the fidelity of
estimation.
\end{abstract}

\vspace{-0.3cm}
\begin{multicols}{2}
\begin{center}
\begin{verse}
{ \small \em
There was an odd qubit from Donegal, \\
who wanted to become most orthogonal. \\
\quad He went through a gate, \\
\quad but not very straight, \\
and came out instead as a Buckyball.
}
\end{verse}
\end{center}
Classical information consists of bits, each of which can be
either $0$ or $1$.  Quantum information, on the other hand,
consists of qubits which are two-level quantum systems with one
level labeled $|0\rangle$ and the other $|1\rangle$.  Qubits can
not only be in one of the two levels, but in any superposition of
them as well.  This fact makes the properties of quantum
information quite different from those of its classical
counterpart.  For example, it is not possible to construct a
device which will perfectly copy an {\it arbitrary} qubit
\cite{Wootters,Buzek1} while the copying of classical information
presents no difficulties. Another difference between classical and
quantum information is as follows: It is not a problem to
complement a classical bit, i.e. to change the value of a bit, a
$0$ to a $1$ and vice versa.  This is accomplished by a NOT gate.
Complementing a qubit, however, is another matter.  The complement
of a qubit $|\Psi\rangle$ is the qubit $|\Psi^{\perp}\rangle$
which is orthogonal to it.  Is it possible to build a device which
will take an {\it arbitrary} (unknown)
 qubit and transform it into the qubit
orthogonal to it?

The best intuition for this problem is obtained by looking at the
desired operation as an operation on the Poincar\'e sphere, which
represents the set of pure states of a qubit system. Thus every
state, pure or mixed, is represented by a vector in a
three-dimensional
space, whose components are the expectations of the three Pauli
matrices. The full state space is thereby mapped onto the unit
ball, whose surface represents the set of pure states. In this
picture the ambiguity of choosing an overall phase for $|\Psi\r$
is already eliminated. The points corresponding to $|\Psi\r$ and
$|\Psi^\perp\r$ are antipodes of each other. The desired
Universal-NOT (U-NOT) operation is therefore nothing but the
{\it inversion of the Poincar\'e sphere}.

Note that the inversion preserves angles (related in a simple way
to the scalar product $\vert\l\Phi,\Psi\r\vert$ of rays), so by
Wigner's Theorem the ideal U-NOT must be implemented either by a
unitary or by an anti-unitary operation. Unitary operations
correspond to proper rotations of the Poincar\'e sphere, whereas
anti-unitary operations correspond to orthogonal transformations
with determinant $-1$. Clearly, the U-NOT operation is of the
latter kind, and an anti-unitary operator $\Theta$ (unique up to a
phase) implementing it is
\begin{equation}
\Theta\bigl(\alpha\vert0\r+ \beta\vert1\r\bigr)
   =\beta^{\ast}\vert0\r-\alpha^{\ast}\vert1\r.
\label{theta}
\end{equation}

The difficulty with anti-unitarily implemented symmetries is that
they are not completely positive, i.e., they cannot be applied to
a small system, leaving the rest of the world alone. (The tensor
product of an anti-linear and a linear operator is ill-defined).
Thus time-reversal, perhaps the best known operation of this kind,
can only be a global symmetry, but makes no sense when applied
only to a subsystem. By definition, a ``gate'' is an operation
applied to only a part of the world, so must be represented by a
completely positive operation. By the Stinespring Dilation Theorem
this is equivalent to saying that any gate must have a realization
by coupling the given system to a larger one (some ancillas),
performing a unitary operation on the large system, and
subsequently restricting to a subsystem. Hence an ideal U-NOT gate
does not exist.

The same is true, of course, for other anti-unitarily implemented
operations like the complex conjugation (or equivalently the
transposition) of the density matrix, which corresponds to the
reflection of the sphere at the $x_2=0$ plane, because only the
Pauli matrix $\sigma_2$ has imaginary entries. Clearly, any such
operation can be represented by a U-NOT, followed by a suitable
unitary rotation, and conversely. On the other hand, if we relax
the ``universality'' condition, the U-NOT operation may become
viable: if we are promised that the elements of the density matrix
(or the components of $|\Psi\r$) are {\it real}, the states lie in
the $x_2=0$ plane so that the inversion at the center is
equivalent to a proper rotation by $\pi$ around the $x_2$-axis.

Because we cannot design a perfect Universal-NOT gate, what we
would like to do is see how close we can come. At this point we
can consider two scenarios. The first one is based on the
measurement of input qubit(s) -- using the results of an optimal
measurement we can manufacture an orthogonal qubit, or any desired
number of them. Obviously, the fidelity of the NOT operation in
this case is equal to the fidelity of estimation of the state of
the input qubit(s). The second scenario would be to approximate an
anti-unitary transformation on a Hilbert space of the input
qubit(s) by a unitary transformation on a larger Hilbert space
which describes the input qubit(s), blank qubits which are to
become the complements, and the quantum device playing the r\^ole
of the gate. We demand that the gate performs equally well for any
(unknown) pure input state, so it is natural to focus on {\it
universal} gates ``U-NOT'', i.e., gates which treat every state
vector in the same way in the sense of unitary symmetry. In what
follows we shall address both scenarios.

In order to state our problem precisely, let $\Hi=\Cx^2$ denote the
two-dimensional Hilbert space of a single qubit
Then the input state of $N$ systems prepared in the
pure state $|\Psi\r$ is the $N$-fold tensor product $|\Psi\r^{\otimes
N}\in\Hi^{\otimes N}$. The corresponding density matrix is
$\rho\equiv \sigma^{\otimes N}$, where $\sigma=\vert\Psi\r\l\Psi\vert$ is the
one-particle density matrix.
An important observation is that the
vectors $|\Psi\r^{\otimes N}$ are invariant under permutations of all
$N$ sites, i.e., they belong to the symmetric, or
``Bose''-subspace $\Hi^{\otimes N}_+\subset\Hi^{\otimes N}$. Thus
as long as we consider only pure input states we can assume all
the input states of the device under consideration to be density
operators on $\Hi^{\otimes N}_+$. We will denote by $\state({\cal
H})$ the density operators over a Hilbert space $\Hi$. Then the
U-NOT gate must be
a completely positive
trace preserving map $T:\state\left(\Hi^{\otimes
N}_+\right)\rightarrow \state(\Hi)$. Our aim is to design $T$ in
such a way that for any pure one-particle state
$\sigma\in\state(\Hi)$ the output $T(\sigma^{\otimes N})$ is as
close as possible to the orthogonal qubit state
$\sigma^\perp=\openone-\sigma$. In
other words, we are trying to make the fidelity $
{\cal F}:=\tr[\sigma^\perp T(\sigma^{\otimes N})]=1-\Delta$
of the optimal complement
with the result of the transformation $T$
 as close as possible to unity for an arbitrary input state. This
corresponds to the problem of finding the minimal value of the error
measure $\Delta(T)$ defined as
\begin{equation}
   \Delta(T)=\max_{\sigma, \rm pure}
   \tr\left[\sigma\ T(\sigma^{\otimes N})\right].
\label{Delta}
\end{equation}

Note that this functional $\Delta$ is completely unbiased with
respect to the choice of input state. More formally, it is
invariant with respect to unitary rotations (basis changes) in
$\Hi$: When $T$ is any admissible map, and $U$ is a unitary on
$\Hi$, the map $T_U(\rho)=U^*T(U\upN\rho{U^*}\upN)U$ is also
admissible, and satisfies $\Delta(T_U)=\Delta(T)$.
We will show later on that one may look for optimal gates $T$,
minimizing $\Delta(T)$, among the {\it universal} ones, i.e., the gates
satisfying $T_U=T$ for all $U$. For such U-NOT gates, the
maximaization can be omitted from the definition (\ref{Delta}),
because the fidelity $\tr\left[\sigma\ T(\sigma^{\otimes
N})\right]$ is independent of $\sigma$.

{\it Measurement-based scenario}\newline
An estimation device by definition takes an input state
$\rho\in\state(\Hi^{\otimes N}_+)$ and produces, on every single
experiment, an ``estimated pure state'' $\sigma\in\state(\Hi)$. As
in any quantum measurement this will not always be the same
$\sigma$, even with the same input state $\rho$, but a random
quantity. The estimation device is therefore described completely
by the probability distribution of pure states it produces for
every given input. Still simpler, we will characterize it by the
corresponding probability density with respect to the unique
normalized measure  on the pure states (denoted ``$d\Phi$'' in
integrals), which is also invariant under unitary rotations. For
an input state $\rho\in\state(\Hi^{\otimes N}_+)$, the value of
this probability density at the pure state $|\Phi\r$ is
\begin{equation}
   p(\Phi,\rho)=
      (N+1)\l\Phi^{\otimes N},\rho\,\Phi^{\otimes N}\r.
\label{pdense}
\end{equation}
To check the normalization,
note that $\int d\Phi\,p(\Phi,\rho)=\tr[X\rho]$ for a suitable
operator $X$, because the integral depends linearly on $\rho$. By
unitary invariance of the measure ``$d\Phi$'' this operator
commutes with all unitaries of the form $U^{\otimes N}$, and since
these operators, restricted to $\Hi\upN_+$ form an irreducible
representation of the unitary group of $\Hi$ [for $d=2$, it is
just the spin $N/2$ irreducible representation of SU(2)], the
operator $X$ is a multiple of the identity. To determine the
factor, one inserts $\rho=\openone$, and uses the normalization of
``$d\Phi$'' to verify that $X=1$.

Note that the density (\ref{pdense}) is proportional to
$\vert\l\Phi,\Psi\r\vert^{2N}$, when $\rho=\vert\Psi^{\otimes
N}\r\l\Psi^{\otimes N}\vert$ is the typical input to such a
device: $N$ systems prepared in the same pure state $|\Psi\r$. In
that case the probability density is clearly peaked sharply at
states $|\Phi\r$ which are equal to $|\Psi\r$ up to a phase.

Suppose now that we combine the state estimation with the
preparation of a new state, which is some function of the
estimated state. The overall result will then be the integral of
the state valued function with respect to the probability
distribution just determined. In the case at hand the desired
function is $f(\Phi)=(\openone-\vert\Phi\r\l\Phi\vert)$. So the
result of the whole measurement-based (``classical'') scheme is
\begin{equation}
\sigma^{(out)}   =  T(\rho)
    =\int d\Phi\ p(\Phi,\rho)\
        \left(\openone-\vert\Phi\r\l\Phi\vert\right).
\label{T-estim}
\end{equation}
The fidelity required for the
computation of $\Delta$ from Eq.(\ref{Delta}) is then equal to
(see also \cite{Derka})
\begin{eqnarray}
\Delta
    =(N+1)\int d\Phi\ \vert\l\Phi,\Psi\r\vert^{2N}
              (1-\vert\l\Phi,\Psi\r\vert^2)
    =\frac{1}{N+2},
\label{Delta-estim}
\end{eqnarray}
where we have used that the two integrals have exactly the same
form (differing only in the choice of $N$), and that the first
integral is just the normalization integral. Since this expression
does not depend on $\sigma$, we can drop the maximization in the
definition (\ref{Delta}) of $\Delta$, and find
$\Delta(T)=1/(N+2)$, from which we find that the fidelity of creation
of a complement to the original state $\sigma$ is ${\cal
F}=\frac{N+1}{N+2}$. Finally we note, that the result of the operation
(\ref{T-estim}) can be expressed in the form
\begin{equation}
\sigma^{(out)}=s_{_N}\sigma^\perp+\frac{1-s_{_N}}{2}
\openone,
\label{8}
\end{equation}
 with the ``scaling'' parameter
$s_{_N}=\frac{N}{N+2}$.
From here it is seen that
 in the limit $N\rightarrow\infty$, perfect estimation of the
input state can be performed, and, consequently, the perfect
complement can be generated. For finite $N$ the mean fidelity is
always smaller than unity. The advantage of the measurement-based
scenario is that once the input qubit(s) is measured and its state
is estimated an arbitrary number $M$ of identical (approximately)
complemented qubits can be produced with the same fidelity, simply
by replacing the output function
$f(\Phi)=(\openone-\vert\Phi\r\l\Phi\vert)$ by
$f_M(\Phi)=(\openone-\vert\Phi\r\l\Phi\vert)^{\otimes M}$.

\bigskip

{\it Quantum scenario: U-NOT gate}\newline

Let us assume we have $N$ input qubits in an unknown state
$|\Psi\r$ and we are looking for a transformation which generates
$M$ qubits at the output in a state as close as possible to the
orthogonal state $|\Psi^{\perp}\r$. The universality of the
proposed transformation has to guarantee that an arbitrary  input
state is complemented  with the same fidelity. If we want to
generate $M$ approximately complemented qubits at the output, the
U-NOT gate has to be represented by $2M$ qubits (irrespective of
the number, $N$, of input qubits), $M$ of which will only serve as
ancilla, and $M$ of which become the output complements. We will
indicate these subsystems by subscripts ``a''=input,
``b''=ancilla, and ``c''=(prospective) output. The U-NOT gate
transformation, ${U}_{NM}$, acts on the tensor product of all
three systems. The gate is always prepared in some state
$|X\r_{bc}$, independently of the input state $|\Psi\r$. The
transformation is determined  by the following explicit expression,
valid for every unit vector $|\Psi\r \in\Hi$:
\end{multicols}
\vspace{-0.5cm}
\noindent\rule{0.5\textwidth}{0.4pt}\rule{0.4pt}{\baselineskip}
\widetext
\be
 {U}_{NM}|N\Psi\r_a\otimes|X\r_{bc}
    =\sum_{j=0}^{M}  \gamma_j |X_j(\Psi)\r_{ab}
      \otimes |\{(M-j)\Psi^\perp;j\Psi\}\r_{c}
~~;\quad
\gamma_j
   = (-1)^j {N+M-j \choose N }^{1/2} { N+M+1 \choose M }^{-1/2},
\label{11}
\ee
\begin{multicols}{2} \noindent
where $|N\Psi\r_{a}=|\Psi\r^{\otimes N}$ is the input state
consisting of $N$ qubits in the same state $|\Psi\r$. On the right
hand side of Eq.(\ref{11}) $|\{(M-j)\Psi^\perp;j\Psi\}\r_{c}$
denotes symmetric and normalized states with $(M-j)$ qubits in the
complemented (orthogonal)  state $|\Psi^\perp\r$ and $j$ qubits in
the original state $|\Psi\r$. Similarly, the vectors
$|X_j(\Psi)\r_{ab}$ consist of $N+M$ qubits, and are given
explicitly by
\be
|X_j(\Psi)\r_{ab} =
|\{(N+M-j)\Psi;j\Psi^{\perp}\}\r_{ab}.
\label{12}
\ee
Here the coefficients $\gamma_j$ were chosen so that the scalar
product of the right hand side with a similar vector written out
for $|\Phi\r$, becomes $\langle \Psi,\Phi\rangle^N$. This implies at the
same time that $U_{NM}$ is linear and that it is unitary after
suitable extension to the orthogonal complement of the vector
$\vert X\rangle_{bc}$.

Each of the $M$ qubits at the output of the U-NOT gate is
described by the density operator (\ref{8}) with
$s_{_N}=\frac{N}{N+2}$, {\em irrespective} of the number of
complements produced. The fidelity of the U-NOT gate depends only
on the number of inputs. This means that this U-NOT gate can be
thought of as producing an approximate complement and then cloning
it, with the quality of the cloning independent of the number of
clones produced. The universality of the transformation is
directly seen from the ``scaled'' form of the output operator
(\ref{8}).

We stress that the fidelity of the U-NOT gate (\ref{11}) is
exactly the same as in the measurement-based scenario. Moreover,
it also behaves as a classical (measurement-based) gate in a sense
that it can generate an arbitrary number of complements with the
same fidelity. We have also checked that these cloned complements
are pairwise separable.

The $N+M$ qubits at the output of the gate which do not represent the
complements are individually in the state described by the density
operator
\be
\sigma^{(out)}_j= s \sigma
+\frac{1-s}{2}\openone ,\qquad j=1,\dots,N+M ~,
\label{15}
\ee
with the scaling factor $s=\frac{N}{N+2}+\frac{2N}{(N+M)(N+2)}$
i.e. these qubits are the {\it clones} of the original state with a fidelity
of cloning larger than the fidelity of estimation. This
fidelity depends on the number, $M$, of clones  produced out of
the $N$ originals,
and in the limit $M\rightarrow\infty$ the fidelity of cloning becomes
equal to the fidelity of estimation.
These qubits
represent the output of the {\it optimal} $N\rightarrow N+M$ cloner
introduced by Gisin and Massar \cite{Gisin}. This means that the
U-NOT gate as presented by the transformation in Eq.\ (\ref{11})
serves also as a universal cloning machine.

At this point the question arises whether the transformation
(\ref{11}) represents the {\it optimal} U-NOT gate via quantum
scenario. If this is so, then it would mean that the
measurement-based and the quantum scenarios realize the U-NOT gate
with the same fidelity.

{\bf Theorem.} {\it Let $\Hi$ be a Hilbert space of dimension
$d=2$. Then among all completely positive trace preserving maps
$T:\state\left(\Hi^{\otimes N}_+\right)\rightarrow \state(\Hi)$,
the measurement-based U-NOT scenario (\ref{T-estim}) attains the
smallest possible value of the error measure defined by
Eq.(\ref{Delta}), namely $\Delta(T)=1/(N+2)$.}

We have already shown [see Eq.(\ref{Delta-estim})] that for the
measurement-based strategy the error $\Delta$ attains the minimal
value $1/(N+2)$. The more difficult part, however, is to show that
no other scheme [i.e., quantum scenario] can do better. Here we
will largely follow the arguments in \cite{WKeyl}.

Recall first the rotation invariance of the functional $\Delta$,
noted after Eq.(\ref{Delta}). Moreover, $\Delta$ is defined as the
maximum of a collection of linear functions in $T$, and is
therefore convex. Putting these observations together we get
\begin{equation}
  \Delta(\hat T)
     \leq\int dU\ \Delta(T_U)
     = \Delta(T),
\label{hatT}
\end{equation}
where $\hat T=\int dU\ T_U$ is the average of the rotated
operators $T_U$ with respect to the Haar measure on the unitary
group. Thus $\hat T$ is at least as good as $T$, and is a {\it
universal} NOT gate ($\hat T_U=\hat T$). Without loss
we  will therefore assume from now on that $T_U=T$ for all $U$.

An advantage of this step is that a very explicit general form for
universal operations is known from the ``covariant form'' of the
Stinespring Dilation Theorem (see \cite{WKeyl} for a version
adapted to our needs). The form of $T$ is further simplified in
our case by the fact that both representations involved are
irreducible: the defining representation of SU(2) on $\Hi$, and
the representation by the operators $U\upN$ restricted to the
symmetric subspace $\Hi\upN_+$. Then $T$ can be represented as a
convex combination $T=\sum_j\lambda_jT_j$, with $\lambda_j\geq0,
\sum_j\lambda_j=1$, and $T_j$ universal gates in their own right,
but of an even simpler form. Universality of $T$ already implies
that the maximum can be omitted from the definition (\ref{Delta})
of $\Delta$, because the fidelity no longer depends on the pure
state chosen. In a convex combination of universal operators $T_j$
we therefore get

\begin{equation}
  \Delta(T)=\sum_j\lambda_j\Delta(T_j).
\label{Delta-aff}
\end{equation}
Minimizing this expression is
obviously equivalent to minimizing with respect to the discrete
parameter $j$.

We write the general form of the extremal gates $T_j$ in
terms of expectation values of the output state for an observable
$X$ on $\Hi$:
\begin{equation}
  \tr\bigl[T(\rho)X\bigr]
     =\tr\bigl[\rho\,V^*(X\otimes\openone)V\bigr],
\label{T-extr} \end{equation} where
$V:\Hi\upN_+\to\Hi\otimes\Cx^{2j+1}$ is an isometry intertwining
the respective representations of SU(2), namely the restriction of
the operators $U\upN$ to $\Hi\upN_+$ (which has spin $N/2$) on the
one hand, and the tensor product of the defining representation
(spin-$1/2$) with the irreducible spin-$j$ representation. By the
triangle inequality for Clebsch-Gordan reduction, this implies
$j=(N/2)\pm(1/2)$, so only two terms appear in the decomposition
of $T$. It remains to compute $\Delta(T_j)$ for these two values.
Omitting the details of the calculations (these follow closely the
arguments presented in Ref.\cite{WKeyl}) we find that
\begin{equation}
  \Delta(T)=\cases{1   & for $j= \frac{N}{2}+ \frac{1}{2}$\cr
              \frac{1}{N+2}  & for $j= \frac{N}{2}- \frac{1}{2}$}.
\label{deltaFin}
\end{equation}
The first value corresponds to getting the state $\sigma$ from a
set of $N$ copies of $\sigma$. The fidelity $1$ is expected for
this trivial task, because taking any one of the copies will do
perfectly. On the other hand, the second value is the minimal
error in the {\it optimal} U-NOT gate, which we were looking for.
This clearly coincides with the value (\ref{Delta-estim}), so the
Theorem is proved.

{\it R\^ole of a priori knowledge}\newline
As was noted earlier, if the input
state $|\Psi\rangle =\alpha |0\rangle +\beta
|1\rangle$ is restricted to the case where the
coefficients $\alpha$ and $\beta$ are real, then
it is possible to construct a perfect quantum NOT gate. A
measurement-based strategy in this case does not do
as well.  Specifically,
 the mean fidelity
of optimal estimation in the present case increases as a function of
input qubits as (see \cite{Derka})
${\cal F}= \frac{1}{2}+\frac{1}{2^{N+1}}\sum_{j=0}^{N-1}
\sqrt{{N\choose j}{N \choose j+1}}$,
and it attains a value equal to unity
only in the limit $N\rightarrow\infty$. This means that with
{\it a priori} knowledge of the set of inputs, the quantum
NOT can perform better
than the measurement-based strategy.

Summarizing our conclusions, we have shown
that there is another difference between
classical and quantum information: classical
bits can be complemented, while arbitrary
qubits cannot.  It is, none the less,
possible to construct approximate quantum-complementing
devices the quality of whose output
is independent of the state of their input. These
devices we called U-NOT gates.  They are closely
related to quantum cloners, and exploiting this
connection it is possible to find an explicit
transformation for a
$N$-qubit input and $M$-qubit output U-NOT gate.
When there is no {\it a priori} information available
 about the state of input qubits  then these U-NOT gates
do not do better than a
measurement-based strategy. On the other hand, as we have shown,
partial {\it a priori} information can dramatically improve performance
of the U-NOT gate.

This work was in part supported by the Royal Society and by
the Slovak Academy of Sciences.
V.B. and R.F.W. thank
the Benasque Center for Physics where part of this work was
carried out.

\end{multicols}

\end{document}